\documentstyle[aps]{revtex}

\newif\ifletter
\newdimen\mathindent
\newcommand{\fl}{\hspace*{-\mathindent}}
\def\ack{\ifletter\bigskip\noindent\ignorespaces\else
    \section*{Acknowledgments}\fi}

\begin{document}

\title{Second order perturbations of a Schwarzschild black hole}

\author{Reinaldo J. Gleiser\dag, Carlos O. Nicasio\dag\ddag, 
Richard H. Price\S, Jorge Pullin\ddag}
\address{\dag\ Facultad de Matem\'atica, Astronom\'{\i}a y F\'{\i}sica,
Universidad Nacional de C\'ordoba,\\ Ciudad
Universitaria, 5000 C\'ordoba, Argentina.}
\address{
\ddag\ Center for Gravitational Physics and Geometry, Department of
Physics,\\
The Pennsylvania State University, 
104 Davey Lab, University Park, PA 16802}
\address{
\S\ Department of Physics, University of Utah, Salt Lake City, Utah
84112.}

\maketitle
\begin{abstract}
We study the even-parity $\ell=2$ perturbations of a Schwarzschild
black hole to second order.  The Einstein equations can be reduced to
a single linear wave equation with a potential and a source term.  The
source term is quadratic in terms of the first order perturbations.
This provides a formalism to address the validity of many first
order calculations of interest in astrophysics.
\end{abstract}

\pacs{4.30+x}

\vspace{-7.5cm} 
\begin{flushright}
\baselineskip=15pt
CGPG-95/10-7  \\
gr-qc/9510049\\
\end{flushright}
\vspace{5.5cm}

Black hole perturbation theory has been a ubiquitous tool in the
analysis of astrophysical situations without symmetries.  It
has played an important role in the study of the expected gravitational
radiation from processes like the infall of matter into a hole
\cite{DRIPP}, and more recently the collision of two black holes
\cite{PrPu}.  
For black holes, perturbations to linear order in a dimensionless
expansion parameter $\epsilon$ can be described through a simple
scalar wave equation in two spacetime dimensions, and yet can describe
complex situations of no particular symmetry.  In contrast, a full
numerical simulation of the nonlinear Einstein equations is at present
prohibitive for most situations of physical interest.

In spite of these advantages there is an important limitation in the
use of linearized theory: there is no information within linearized
theory to determine its range of applicability, i.e., to determine
what values of $\epsilon$ are sufficiently small.  Largely due to this
last problem, linearized theory has usually been limited in use to
situations departing only very slightly from an undisturbed black
hole. However, recent comparisons between linearized theory and full
numerical simulations for the head-on collisions of two black holes
\cite{PrPu} have shown that linearized theory can work remarkably well
in domains in which it would be supposed to fail. The qualitative
explanation for this is that strong nonlinearities near the horizon
can be absorbed by the black hole and therefore not affect the outgoing
radiation.

In order to promote linearized theory to a serious tool for predicting
astrophysical answers and in particular to provide benchmarks for
difficult full numerical simulations, we need a reliable measure of
the errors in a perturbation result for a given value of $\epsilon$.
One can try to construct simply implemented analytic {\em a priori}
indices of the validity of perturbation theory, e.g., the violation of
the exact Hamiltonian constraint by the linearized data\cite{suen}, or
the violation of the linearized Hamiltonian constraint by the exact
initial data\cite{aa}. Such measures are interesting because of their
simplicity, but are not unique and worse, they are not guaranteed to
work in all situations. For instance, their success for head-on
collisions of two black holes gives no reliable information about
their value for less symmetric collisions.

There is, in fact, only one generally reliable index of the accuracy
of linearized perturbation theory: to calculate answers to the next
order in $\epsilon$. Where the higher-order results and the linearized
results differ by (say) 10\% is a point at which one has some
confidence that either  answer is accurate to within around 10\%.
Higher order perturbation calculations seem at the outset to be an
obvious extension of the familiar techniques of linearized
perturbation theory, with a guarantee of simple
results. It turns out --- and this is one of the major points we wish
to make --- that there are some subtle issues of gauge that must be
recognized. Despite these issues, and despite some necessarily lengthy
calculations, second order computations are far more easily done than
the development of full numerical relativity codes. And the difficulty
of the higher order calculations is a necessary price well worth
paying for the advantages provided by reliable perturbation theory
results.


The basic structure of higher order perturbation theory starts with
the same basic formalism as linearized theory.  We write the metric as
$g_{\mu\nu}^{(0)}+ \epsilon \,g_{\mu\nu}^{(1)}
+\epsilon^2\,g_{\mu\nu}^{(2)}+{\cal O}(\epsilon^3) $ where
$g_{\mu\nu}^{(0)}$, the ``background,'' is a known solution of the
vacuum field equations. We write the sourceless Einstein equations as
\begin{equation}\label{formofG}
{\cal G}(g_{\mu\nu}^{(0)}+
\epsilon \,g_{\mu\nu}^{(1)}
+\epsilon^2\,g_{\mu\nu}^{(2)})=0\,,
\end{equation}
where ${\cal G}$ represents the actions of taking partial derivatives
and algebraic combinations to form the components of the Einstein
tensor.  If we expand (\ref{formofG}) in $\epsilon$, the term of order
$\epsilon^0$ automatically vanishes if $g_{\mu\nu}^{(0)}$ is a
background solution. The terms first order in $\epsilon$ can be
written in the form
\begin{equation}
\label{linearG}
\epsilon\,{\cal L}( g_{\mu\nu}^{(1)})=0\,,
\end{equation}
where ${\cal L}$ is a set of differentiations and combinations with
details that depends on $g_{\mu\nu}^{(0)}$. These operations are all
linear in $g_{\mu\nu}^{(1)}$, and eqs.~(\ref{linearG}) 
constitute linearized perturbation theory, a reduction of the problem
to a system of linearized equations.  If the background is the
Schwarzschild geometry, it is well known that these linear equations
can be decomposed into multipole moments, can be separated into two
sets of independent functions (the even-parity and odd-parity
perturbations), and can be rearranged into two simple linear wave
equations in the variables $r,t$: the Regge-Wheeler\cite{rw} equation
for the odd-parity perturbations, and the Zerilli\cite{zerilli}
equation for even-parity.

The part of
(\ref{formofG}) that is proportional to $\epsilon^2$ has two kinds of
terms. There are terms that are linear in $g_{\mu\nu}^{(2)}$, and
terms that are quadratic in $g_{\mu\nu}^{(1)}$. It is clear that the
former terms occur in precisely the same form as do the
$g_{\mu\nu}^{(1)}$ terms in (\ref{linearG}). The set of $\epsilon^2$
terms can then be written as
\begin{equation}
\label{quadG}
\epsilon^2\,{\cal L}( g_{\mu\nu}^{(2)})= \epsilon^2\,{\cal J}(
g_{\mu\nu}^{(1)}, g_{\mu\nu}^{(1)})\,,
\end{equation}
where ${\cal J}$ is quadratic in the first order perturbations. In
solving for the second order perturbations, one treats the first order
perturbations as already known, so ${\cal J}$ plays the role of a
source term in (\ref{quadG}).

The ${\cal L}$ operator in (\ref{quadG}) is precisely the same
operator as in (\ref{linearG}), so for each linearized theory equation
for $g_{\mu\nu}^{(1)}$ there is a corresponding equation for
$g_{\mu\nu}^{(2)}$, differing only in the presence of a ``source''
term. The very same manipulations that lead to the Regge-Wheeler and
Zerilli equations, must therefore lead to Regge-Wheeler and Zerilli
equations of precisely the same form for second order perturbations,
except for the presence of source terms.  We therefore have a
guarantee at the outset that, as in first order theory, we can derive
a simple wave equation for the perturbations.

The form of the metric functions, to any order in $\epsilon$, depend
on how the spacetime coordinates are defined, to that order in
$\epsilon$. We assume that we have a set of coordinates $t,r,\theta,\phi$
in terms of which the metric, to zero order in $\epsilon$, takes on
the standard Schwarzschild form. The background
metric $g_{\mu\nu}^{(0)}$ then is the Schwarzschild metric. 
We can transform to new coordinates ${x^{\alpha}}'$ with
$\epsilon$-dependent coordinate transformations
\begin{equation}
\label{coordxfrm}
{x^{\alpha}}'= {x^{\alpha}}+\epsilon X^{\alpha}_{(1)}(x^{\beta})
+\epsilon^2X^{\alpha}_{(2)}(x^{\beta})+{\cal O}(\epsilon^3)\ .
\end{equation}
The form of the background metric $g_{\mu\nu}^{(0)}$ is unchanged by
the transformation, and we call such a  coordinate change a gauge
transformation. If the first order perturbation functions
$g_{\mu\nu}^{(1)}$ are changed by the coordinate change (i.e., if
$X^{\alpha}_{(1)}(x^{\beta})\neq0$) we call it a first order gauge
transformation. Note that a first order gauge transformation will in
general also change the form of the second order perturbations
$g_{\mu\nu}^{(2)}$. We can also make gauge transformations which
leave $g_{\mu\nu}^{(1)}$ invariant and which change
$g_{\mu\nu}^{(2)}$ (i.e., transformations with
$X^{\alpha}_{(1)}(x^{\beta})=0$, but for which
$X^{\alpha}_{(2)}(x^{\beta})\neq0$).  These we call second order gauge
transformations.

We will want to take advantage of the freedom to choose coordinates in
order to impose some simplifying special conditions on the metric
perturbations.  As an example, for even parity perturbations we will
at one stage be setting $g_{r\theta}=0$ through second order in
$\epsilon$. To do this one makes a first order gauge transformation to
set $g_{r\theta}^{(1)}=0$. This transformation will affect the second
order perturbations in some way that depends on the details of the
coordinate transformation. One next makes a second order
transformation to set $g_{r\theta}^{(2)}=0$.  This leaves
$g_{r\theta}^{(1)}=0$, since a second order transformation cannot
change the first order perturbations.

In higher order perturbation calculations new questions can arise,
such as what it means for a quantity to be gauge invariant.  It turns
out to be useful always to think of coordinate transformation as a
sequence of distinct steps, as in the example above. In this way we
see that a quantity can be ``second order gauge invariant'' (invariant
only under a second order transformation), ``first order invariant''
(invariant only to first order under a first order transformation) or
``first and second order invariant'' (invariant up to second order for
a general transformation (\ref{coordxfrm})).


In this paper we take the first important steps in using second order
perturbation theory as a tool; we provide the formalism for
calculating the second order contribution to outgoing radiation.  We
will, however, make important, and yet physically sensible
restrictions in our formulation.  First, we will restrict attention to
the even-parity $\ell=2, m=0$ second order contributions. This is
justified by the fact that most radiation processes are dominated by
quadrupolar radiation, and we are primarily interested in the ``error
measure'' on the first order quadrupole calculation. We give here the
even-parity formalism, but the odd-parity equivalent follows the same
pattern and is in fact simpler. The restriction to $m=0$ is for
simplicity only, the generalization to $m\neq0$ is immediate. Our
second, and more subtle, restriction is on the contributions to the
``source'' term.  Since the source is quadratic in the first order
perturbations, there is an infinite number of first order multipoles
contributing to the $\ell=2$ projection of the source. We restrict
attention only to the first order $\ell=2$ contributions. There are
several reasons for this: (i) It is primarily a simplifying
assumption; the $\ell=2$ projection of the {\it complete} source is
straightforward to compute either numerically or as an infinite
series, but would introduce unnecessarily distracting complications
here. (ii) As a practical matter, it is plausible that in most 
practical situations  the source will be dominated by the first
order quadrupole terms. (iii)
The specific case to which we will first apply this formalism (details
to be published elsewhere) is the head-on collision from initially
small separation\cite{PrPu}, in which the initial separation is the
expansion parameter. In that case it can be shown that the first order
perturbations have {\em only } an $\ell=2$ contribution.

In terms of the usual Schwarzschild-like coordinates $t,r,\theta,
\phi$, and following the Regge-Wheeler\cite{rw} prescription and
notation, we write the general form of the  $\ell =2$, even parity,
perturbation of the spherically symmetric black hole metric in the
form,
\begin{eqnarray}\fl
g _{tt}&=&-(1-2M/r) \left\{1 - \left[\epsilon
  {H}^{(1)}_0+
\epsilon^2   {H}^{(2)}_0\right] P_2(\theta) \right\} \nonumber \\\fl
g _{rr}&=&(1-2M/r)^{-1} \left\{1 +\left[\epsilon
  {H}^{(1)}_2+\epsilon^2
  {H}^{(2)}_2 \right] P_2(\theta)\right\} \nonumber \\\fl
g _{rt}&=&\left[\epsilon   {H}^{(1)}_1+\epsilon^2   {H}^{(2)}_1 \right]  
P_2(\theta) \nonumber \\\fl
g _{t \theta}&=& \left[\epsilon   {h}^{(1)}_0 + \epsilon^2
{h}^{(2)}_0 \right]  P'_2(\theta) \nonumber  \\\fl
g _{r \theta}&=& \left[ \epsilon   {h}^{(1)}_1 + \epsilon^2
{h}^{(2)}_1 \right]
  P'_2(\theta) \nonumber  \\\fl
g _{\theta\theta}&=& r^2 \left\{1+[\epsilon
  {K}^{(1)} +\epsilon^2  {K}^{(2)} ] P_2(\theta)+ [\epsilon
  {G}^{(1)}  + \epsilon^2   {G}^{(2)} ]
P''_2(\theta) \right\} \nonumber \\\fl
\label{pertmet}
g _{\phi\phi} &=&r^2 \left\{1+ \sin^2\theta
[\epsilon   {K}^{(1)}  + \epsilon^2   {K}^{(2)} ] P_2(\theta)
+\sin(\theta)
\cos(\theta)[ \epsilon   {G}^{(1)}  +\epsilon^2   {G}^{(2)} ] 
  P'_2(\theta) \right\}
\end{eqnarray}
where $P_2(\theta)=3 (\cos^2\theta-1)/2$, $P'_2(\theta) =\partial
P_2(\theta) /\partial \theta$, and $P''_2(\theta) =\partial^2
P_2(\theta) /\partial \theta^2$, and where the functions $H, h, K$ and
$G$ depend on $t$ and $r$ only.  Just as in \cite{rw}, which
was restricted to linearized theory, we may impose the Regge-Wheeler
gauge conditions through second order, $h^{(i)}_0 = h^{(i)}_1=
G^{(i)}= 0$, $i=1,2$, by the two-step process described above. One can
show, (details will be given in a lengthier paper), that given an
arbitrary perturbation of the form (\ref{pertmet}), there exists
always (locally) a uniquely defined gauge transformation that takes
the metric to the Regge-Wheeler gauge. (One can, in fact, write these
Regge-Wheeler functions, both to first and second order, explicitly in
terms of perturbations in an arbitrary gauge, and through these
expressions view the Regge-Wheeler perturbations as gauge invariant.)

With the linearized equations in the Regge-Wheeler gauge,
Zerilli\cite{zerilli} assumes time dependence $e^{i\omega t}$ (i.e.,
makes a fourier transform), and works with functions of $\omega$ and
$r$, but the process of deriving a single wave equation can be done
equally well in terms of the original $r,t$ variables. One can repeat
this Zerilli process, step by step, with the second order equations,
in the Regge-Wheeler gauge, in which the only new feature is the
inclusion of quadratic ``source'' terms. One finds a total of seven
nontrivial Einstein equations for the four second order Regge-Wheeler
perturbation functions $H_0^{(2)},H_1^{(2)},H_2^{(2)},
K^{(2)}$. These equations are linear in the second order functions,
but quadratic in the first order perturbations. One of these takes the
form
\begin{equation}
{H}^{(2)}_2 = {H}^{(2)}_0 +[ ({H}^{(1)}_1)^2 - ({H}^{(1)}_0)^2]/7
\end{equation}
and may be used to eliminate ${H}^{(2)}_2$ from the other equations.
(We also have ${H}^{(1)}_2 = {H}^{(1)}_0$, from the
first order equations). We are therefore left with six equations for the
three remaining second order functions, but it turns out that we may
choose three of these equations so that they can be obtained by
compatibility of the other three. Moreover, from one of Einstein's
equations we find,
\begin{equation}\label{H2t}
\fl{H}^{(2)}_0,_t = - K^{(2)},_t +[(1-2M/r) H^{(2)}_1],_r
+(1/14)[2(K^{(1)})^2+3(H^{(1)}_0)^2 -3 (H^{(1)}_1)^2],_t
\end{equation}
(a comma indicates partial derivative), which may be used to eliminate
${H}^{(2)}_0$ (or rather its time derivative) in the remaining two
independent equations. Following Zerilli\cite{zerilli}, we introduce
now a pair of functions $\chi(t,r)$ and ${\cal R}(t,r)$, related to
$K^{(2)}(t,r)$, and $H^{(2)}_1(t,r)$ by
\begin{equation}
K^{(2)},_t = {6(r^2+Mr +M^2) \over r^2(2r+3M) } \chi + {\cal R} 
\label{K2t}
\end{equation}
\begin{equation}
H^{(2)}_1 =  { 2r^2-6Mr-3M^2 \over (r-2M)(2r+3M) } \chi + 
{r^2 \over r-2M}{\cal R}
\label{H21}
\end{equation}
By substitution of these relations in Einstein's equations we find,
\begin{eqnarray}\label{Req}
\fl{\cal R} &=& \left( 1 -2 {M \over r}\right) \chi,_r + {1 \over 7}
{r-2M \over 2r+3M} \left( H^{(1)}_0 H^{(1)}_0,_t -2 K^{(1)}
H^{(1)}_0,_t -r K^{(1)},_r H^{(1)}_0,_t \right.\nonumber \\\fl
&&
-2 K^{(1)} K^{(1)},_t - {4
\over r} K^{(1)} H^{(1)}_1 + {r^2 \over r-2M} K^{(1)},_t H^{(1)}_1,_t
 - {2 M \over r-2M} H^{(1)}_0 K^{(1)},_t \nonumber \\\fl
&&\left.+{2 M \over r} H^{(1)}_1
 K^{(1)}_r - H^{(1)}_1 H^{(1)}_1,_t - 2 {r-2M \over r^2} H^{(1)}_0
H^{(1)}_1 \right)
\end{eqnarray}
while the function $\chi(t,r)$ satisfies the single second order
differential equation,
\begin{equation}
{\partial^2 \chi(t,r) \over \partial t^2} -
{\partial^2 \chi(t,r) \over \partial {r^*}^2} 
+V(r)  \chi(t,r) = {\cal S}_{\rm RW}.\label{zerieq}
\end{equation}
where $r^*=r+2 M \log(r/2M-1)$ is the ``tortoise'' radial coordinate,
and where $V(r)$ is the $\ell=2$ Zerilli potential, identical to that
for the linearized theory (see \cite{zerilli}). The ``source term''
${\cal S}_{\rm RW}$ is a quadratic expression in the first order
perturbations.

As guaranteed by the procedure, we have arrived at a Zerilli equation
for a second order Zerilli function, differing from the Zerilli
equation of linearized theory only in the presence of the source term
${\cal S}_{\rm RW}$ quadratic in the first order perturbations. (The
explicit expression for ${\cal S}_{\rm RW}$ is somewhat lengthy and
will be given elsewhere.)  The second order Zerilli equation
(\ref{zerieq}) is then the core of second order perturbation theory.
We have verified, as a consistency check, that these functions satisfy
the second order Einstein equations, (or rather, their time
derivatives \cite{tderiv}), provided that the first order Einstein
equations and (\ref{zerieq}) are satisfied.   In principle,
one first solves the first order Zerilli equation, finds the first
order perturbations, constructs the source term ${\cal S}_{\rm RW}$,
and solves (\ref{zerieq}) for $\chi$.  In practice, however, there are
complications. These are evident in the fact that the source term in
(\ref{zerieq}) diverges at large radius. This divergence does not
indicate singular physical behavior; the second order radiation could
in principle be extracted from $\chi$ computed in this way, but the
procedure would, at best, be computationally inefficient.
In connection with this, two issues concerning gauge choice
should, in particular, be noted.  First, it must be realized that most
of the development of the second order Zerilli equation (\ref{zerieq})
did not require that the gauge choice be Regge-Wheeler for the first
order perturbations; only second order Regge-Wheeler restrictions are
really needed. If we had made another gauge choice, or no gauge
restrictions at all, only the quadratic terms in (\ref{H2t}),
(\ref{Req}), and (\ref{zerieq}) would change. The ``second order
Zerilli equation with source'' is, therefore, not unique. Another
viewpoint on this is that we could start with (\ref{zerieq}) and
introduce a new Zerilli function through 
\begin{equation}\label{newchi}
\chi=\chi_{\rm new}+{\rm quad}
\end{equation}
where quad is any expression quadratic in the first order
perturbations. The new Zerilli function $\chi_{\rm new}$ would then
satisfy (\ref{zerieq}), but with a modified source term. The choice of
the Regge-Wheeler first order gauge is made for
convenience. Expressions in this gauge usually turn out  to be most
compact.

The second gauge issue of note concerns the computation of radiation.
To first order all information about radiation is carried in a
``Zerilli function'' $\psi$ which we define, in the Regge-Wheeler gauge, with 
the notation introduced in (\ref{pertmet}), by
\begin{equation}\label{ourpsi}
\psi= {r (r-2 M)\over 3 (2 r + 3 M)} \left(
H_0^{(1)} -r {\partial K^{(1)} \over \partial r}\right)+
{r \over 3} K^{(1)}.
\end{equation}
Several different normalizations for the ``Zerilli function'' have
been used in the literature, so it is important to specify the
relationship of our choice to others.  If there are no (first order)
sources, the definition in (\ref{ourpsi}) can be shown, for $\ell=2$,
to have the same appearance as the definition of the wave funtion
$\widehat{R}_{LM}$ of Zerilli. That is, (\ref{ourpsi}) specifies the
same combination of perturbation functions $K^{(1)}$ and so forth.
But in our definition (\ref{pertmet}) of the metric perturbation
functions, we have expanded in Legendre polynomials, whereas Zerilli
does his multipole expansions in terms of spherical harmonics. As a
result the actual relationship between our $\psi$ and Zerilli's
$\widehat{R}$ is
\begin{equation}
\psi=\sqrt{\frac{2\ell+1}{4\pi}}\widehat{R}
\end{equation}
Another normalization that is of importance is the normalization used
by Cunningham et al.\cite{cpm}, which agrees with that of Zerilli.
The normalization used in \cite{PrPu} and \cite{anninosetal}, called
$\psi_{\rm pert}$ in the latter reference, is related to $\psi$ used
here by:
\begin{equation}
\psi=\sqrt{\frac{2\ell+1}{4\pi}}\,2\,\frac{(\ell-2)!}
{(\ell+2)!}\psi_{\rm pert}\ .
\end{equation}

The computation of graviational radiation power is done in a
coordinate system that is asymptotically flat. In this system, which
we will denote with tildes, radiation information is carried by the
perturbations in $g_{\theta\theta},g_{\phi\phi}$ and
$g_{\theta\theta}$.  The function $\widetilde{G}$, in the notation of
(\ref{pertmet}), falls off in this gauge as $1/r$, to all orders in
$\epsilon$, and $r\widetilde{G}$ can be thought of as the amplitude of
the even parity gravitational wave.  
In terms of our definitions, The first order part
$\widetilde{G}^{(1)}$ is given by $\widetilde{G},_t^{(1)} =
\psi(t,r),_t/r +O(1/r^2)$ and the first order radiated power can 
be shown to be
\begin{equation}
{\rm Power}=\frac{1}{16}\,
\frac{(\ell+2)!}{(\ell-2)!}\,\frac{1}{2\ell+1}(\epsilon\dot{\psi})^2\ ,
\end{equation}
where $\dot{\psi}$ is the time derivative of $\psi$.
For $\ell=2$ the result is Power $=(3/10)(\epsilon\dot{\psi})^2$.
If we want the gravitational wave amplitude to second order we must
calculate $\epsilon\widetilde{G}^{(1)}
+\epsilon^2\widetilde{G}^{(2)}$.

The computation of $\widetilde{G}^{(2)}$ can be approached in several
ways. One could, in principle, transform to a gauge which is
asymptotically flat to first order, so that the source
term ${\cal
S}_{\rm RW}$ is replaced by the appropriately modified source term
${\cal S}_{\rm rad}$. One then solves for $\chi$, from it computes the
second order Regge-Wheeler metric perturbations, and then does a
second order gauge transformation to a second order asymptotically
flat gauge. In this gauge one reads off $\widetilde{G}^{(2)}$.  In
practice the same thing can be accomplished more conveniently with a
transformation of the form (\ref{newchi}).  We start with
(\ref{zerieq}) and with a gauge which has Regge-Wheeler restrictions
to first and second order, and we introduce a new function
$\chi(t,r)_{\rm ren}$, a sort of ``renormalized $\chi$,'' given by
\begin{equation}\label{chiren}
\chi_{\rm ren} = \chi- {2 \over 7}\left[
{ r^2\over (2 r+3 M)} K^{(1)} K^{(1)},_t+(K^{(1)})^2\right]\ .
\end{equation}
Simple replacement in (\ref{zerieq}) gives us an equation for 
$\chi(t,r)_{\rm ren}$ with the same form as (\ref{zerieq}), but with 
a source term
\begin{eqnarray}\label{rensource}
\fl&&{\cal S}_{\rm ren}  =  {12 \over 7} {\mu^3 \over \lambda} \left[ -{12
(r^2+Mr+M^2)^2 \over r^4\mu^2\lambda} \left(\psi,_t\right)^2 -4
{(2r^3+4r^2M+9rM^2+6M^3) \over r^6\lambda} \psi \psi,_{rr}  \right.
\nonumber \\\fl
& &  +{(112r^5+480r^4M+692r^3M^2+762r^2M^3+441rM^4+144M^5) \over
r^5\mu^2\lambda^3} \psi \psi,_t - {1 \over 3r^2} \psi,_t \psi,_{rrr}
\nonumber \\\fl
& &   +{18r^3-4r^2M-33rM^2-48M^3 \over 3 r^4\mu^2\lambda} \psi,_r
\psi,_t + {12r^3+36r^2M+59rM^2+90M^3 \over 3 r^6\mu} \left(\psi,_r
\right)^2 \nonumber \\\fl
& &\! +\! 12 {(2r^5 + 9r^4M +6r^3M^2\!-\!2r^2M^3\!-\!15rM^4-15M^5) \over
r^8\mu^2\lambda} \psi^2 \!-\!4 {(r^2+rM+M^2) \over r^3\mu^2} \psi,_t\!
\psi,_{tr} \nonumber \\\fl
& & -2 {(32r^5+88r^4M+296r^3M^2+510r^2M^3+561rM^4+270M^5) \over
r^7\mu\lambda^2} \psi \psi,_r  + { 1 \over 3r^2 } \psi,_r \psi,_{trr}
\nonumber \\\fl
& & - {2r^2-M^2 \over r^3\mu\lambda} \psi,_t \psi,_{rr}
 +{8r^2+12rM+7M^2 \over r^4\mu\lambda} \psi \psi,_{tr}  +{3r-7M \over
 3r^3\mu} \psi,_r \psi,_{tr} - {M \over r^3\lambda} \psi \psi,_{trr}
 \nonumber \\\fl
& &  + {4(3r^2+5rM+6M^2) \over 3r^5} \psi,_r \psi,_{rr} \left.  +
{\mu\lambda \over 3 r^4} \left( \psi,_{rr} \right)^2 - {2r+3M \over
3r^2\mu} \left( \psi,_{tr} \right)^2 \right]
\end{eqnarray}
where $\lambda=(2 r+3 M)$, where $\mu=(r-2 M)$, and where $\psi$ is the
solution of the first order Zerilli equation.  This ``renormalized''
source dies off asymptotically for large $r$, and we find that 
$\chi_{\rm ren}$
behaves asymptotically as a function of $t-r^*$ only. We will postpone
a detailed proof to a lengthier publication.

Most important, it can be shown that $\chi_{\rm ren}$ and the radiation
are related by
\begin{equation}
\widetilde{G}^{(2)},_t = 
{1 \over r}
\left[
\chi_{ \rm ren} +
{1 \over 7} {\partial \over \partial t} \left( 
\psi {\partial \psi \over \partial t}\right)\right]+
O(1/r^2)
\label{12}
\end{equation}
and we see that $\chi_{ \rm ren}$ determine completely, (and in a
numerically meaningful way), the asymptotic behavior of
$\widetilde{G}^{(2)},_t$. It seem appropriate, then, to refer to
$\chi_{ \rm ren}$ as the second order Zerilli function.

To conclude, we present the expression of the gravitational radiation
power, accurate to second order in $\epsilon$, from which the total
radiated energy is easily obtained, in terms of the first and
second order Zerilli functions.  To compute it, we just write the
Landau-Lifshitz pseudotensor in terms of the metric perturbations in
the asymptotically flat gauge (see \cite{cpm} for details) and
substitute the expression for the manifestly asymptotically flat form
of the metric perturbation (\ref{12}). The result is
\begin{equation}\label{power}
{\rm Power}= 
{3\over 10} \left\{ \epsilon {\partial \psi \over \partial t} +
\epsilon^2 
\left[
 \chi_{ \rm ren}  +
{1 \over 7} {\partial \over \partial t} \left( \psi 
{\partial \psi \over \partial
t}\right)\right]\right\}^2\ .
\end{equation}

The formalism, finally, consists of the following steps: (i) From a
solution to the initial value problem, one extracts first order
perturbations of three geometry and extrinsic curvature, and from them
computes Cauchy data for the first order Zerilli function $\psi$
\cite{aa}.  (ii) The Zerilli equation is then solved numerically for
$\psi$ in the $t,r$ region of interest.  (iii) The solution for $\psi$
is used to compute ${\cal S}_{\rm ren}$ in (\ref{rensource}). (iv)
From the initial value solution one next uses the definition of $\chi$
in the Regge-Wheeler gauge
\begin{equation}
\chi(t,r)=\frac{r}{2r+3m}\left[
rK^{(2)}_{,t}-\left(1-\frac{2M}{r}
\right)H_1^{(2)}
\right]\,,
\end{equation}
from
(\ref{K2t}) and
(\ref{H21}), and the other second order Einstein equations, at the
initial hypersurface, to find the initial value of $\chi$, and its
time derivative. (v) Next (\ref{chiren}) and its time derivative are
used to find the initial $\chi_{\rm ren}$ and $\dot{\chi}_{\rm ren}$
(vi) The renormalized Zerilli equation [eq.~(\ref{zerieq}) with the
source ${\cal S}_{\rm ren}$] is then solved numerically for $\chi_{\rm
ren}$. (vi) Finally, the outgoing radiation power is computed from
(\ref{power}).  We will be pursuing several applications of this
formalism in subsequent publication.

\ack
This work was supported in part by grants 
NSF-INT-9512894, NSF-PHY-9423950, NSF-PHY-9507719,  by
funds of the University of C\'ordoba, the University of Utah, the
Pennsylvania State University and its Office for Minority Faculty
Development, and the Eberly Family Research Fund at Penn State. We
also acknowledge support of CONICET and CONICOR (Argentina). JP also
acknowledges support from the Alfred P. Sloan Foundation through an
Alfred P. Sloan fellowship.



\begin{thebibliography}{c}
\bibitem{tderiv} This is because we only have expressions for
$K^{(2)},_t$ and $H_0^{(2)},_t$. Since we are solving the vacuum
Einstein equations, this only implies that we should also give
appropriate initial data for these functions. In practical calculations
this is automatically assured when we impose that the data satisfies
the initial value constraints of general relativity.
\end{thebibliography}
\end{document}